\documentclass[12pt,twoside]{article}
\usepackage{amsmath,mathrsfs,bm,amssymb,color}%suzuki
\newtheorem{theorem}{Theorem}[section]
\newtheorem{proposition}[theorem]{Proposition}
\newtheorem{lemma}[theorem]{Lemma}
\newtheorem{corollary}[theorem]{Corollary}

\newtheorem{remark}[theorem]{Remark}

\def\bbbone{{\mathchoice {\rm 1\mskip-4mu l} {\rm 1\mskip-4mu l}
{\rm 1\mskip-4.5mu l} {\rm 1\mskip-5mu l}}}
\def\one{\bbbone}
\newcommand{\cn}{{\ms N}}

\providecommand{\seq}[1]{(#1_n)_{n\in \mathbb{N}}}
\newcommand{\pro}[1]{(#1_t)_{t\in\RR}}
\newcommand{\proo}[1]{(#1_t)_{t\geq0}}
\newcommand{\II}{W_\infty}
\newcommand{\III}{W_\p^\infty}

\newcommand{\proof}{{\noindent \it Proof:\ }}
\newcommand{\qed}{\hfill $\Box$\par\medskip}

\newcommand{\bi}{
\begin{itemize}
}
\newcommand{\ei}{\end{itemize}}
\newcommand{\OF}{\Omega_\fff}

\renewcommand{\d}{\displaystyle}

\newcommand{\BR}
{{{\mathbb R}^d}}
\newcommand{\RR}{{{\mathbb R}}}
\newcommand{\CC}{{{\mathbb C}}}
\newcommand{\NN}{{{\mathbb N}}}
\newcommand{\ZZ}{{{\mathbb Z}}}

\newcommand{\om}{\omega}

\newcommand{\g}{g^2}%\frac{g^2}{2}}

\newcommand{\LR}{{L^2({\mathbb R}^d)}}

\newcommand{\Ebb}{{\mathbb E}}

\newcommand{\fff}{{\ms F}}

\newcommand{\f}{^{-1}}

\newcommand{\BERN}{\ms B}
\newcommand{\bern}{\Psi}

\newcommand{\add}{a^{\ast}}
\newcommand{\p}{{\rm pol}}

\newcommand{\lk}{\left(}
\newcommand{\rk}{\right)}
\newcommand{\lkk}{\left\{}
\newcommand{\rkk}{\right\}}

\newcommand{\TT}{\frac{\alpha}{2}}

\newcommand{\half}{\frac{1}{2}}

\newcommand{\hhh}{{\ms H}}
\newcommand{\ms}{\mathscr}

\newcommand{\hf}{{H_{\rm f}}}

\newcommand{\hn}{H}

\newcommand{\vp}{\hat{\varphi}}
\newcommand{\gr}{{\varphi_{\rm g}}}
\newcommand{\grp}{\varphi_{\rm p}}
\newcommand{\grt}{\varphi_{\rm g}^T}

\newcommand {\n}  {\ensuremath {\mathsf{N}}}

\newcommand{\eq}[1]{\begin{equation}
\label{#1}}
\newcommand{\en}{\end{equation}}

\newcommand{\eqn}
{\begin{eqnarray*}
}
\newcommand{\enn}{\end{eqnarray*}}

\newcommand{\bt}[1]{\begin{theorem}
\label{#1}}
\newcommand{\et}{\end{theorem}}
\newcommand{\bl}[1]{\begin{lemma}
\label{#1}}
\newcommand{\enl}{\end{lemma}}

\newcommand{\bc}[1]{\begin{corollary}
\label{#1}}
\newcommand{\ec}{\end{corollary}}

\newcommand{\kak}[1]{(\ref{#1})}
\newcommand{\la}{\lambda}

\newcommand{\hp}{H_{\rm p}}
\newcommand{\Hp}{H_{\rm pol}}

\makeatletter
\@addtoreset{equation}{section}
\makeatother

\begin{document}
\title
{\large \sc
A Probabilistic Representation of the Ground State Expectation of Fractional Powers of the
Boson Number Operator}
%in the Nelson model}

\author{
\small Fumio Hiroshima\\
{\small\it Faculty of Mathematics, Kyushu University}    \\[-0.7ex]
{\small\it 744 Motooka, Fukuoka, 819-0395, Japan}      \\[-0.7ex]
{\small  {\tt hiroshima@math.kyushu-u.ac.jp}}\\[0.3cm]
\small J\'ozsef L\H{o}rinczi \\
{\small\it School of Mathematics, Loughborough University}    \\[-0.7ex]
{\small\it Loughborough LE11 3TU, United Kingdom}      \\[-0.7ex]
{\small  {\tt J.Lorinczi@lboro.ac.uk}  }\\[0.3cm]
\small Toshimitsu Takaesu \\
{\it \small Faculty of Mathematics, Kyushu University} \\[-0.7ex]
{\it \small 744 Motooka, Fukuoka, 819-0395, Japan} \\[-0.7ex]
{\small {\tt t-takaesu@math.kyushu-u.ac.jp }} \\[-0.7ex]}

\date{}
\maketitle


\begin{abstract}
\noindent
We give a formula in terms of a joint Gibbs measure on Brownian paths and the measure of a random-time Poisson
process of the ground state expectations of fractional (in fact, any real) powers of the boson number operator
in the Nelson model. We use this representation to obtain tight two-sided bounds. As applications, we discuss 
the polaron and translation invariant Nelson models. 
\end{abstract}
\newpage
\setlength{\baselineskip}{14pt}

\section{Introduction}
\subsection{Fractional boson number operator}
Functional integration is a powerful tool in studying spectral properties of self-adjoint operators. In
particular, quantum field theory provides a large number of interesting problems which can be addressed
by using functional integration, see \cite{bhlms} and \cite[Chapter 6]{LHB11}. In a more recent development
we have extended this method to operators originating from relativistic quantum theory such as the
Schr\"odinger operator $\sqrt{-\Delta+m^2}-m+V$, $m\geq0$, fractional Schr\"odinger operators of the
type $(-\Delta)^{\alpha/2} + V$, $0 < \alpha < 2$, and more generally, Bernstein functions of the
Laplacian \cite{hil}. While in the usual case of the Laplacian functional integration is performed over
the paths of Brownian motion, in these cases we have $\alpha$-stable processes and more general subordinate
Brownian motion.

Inspired by this development, in this paper we are concerned with fractional and general non-integer powers
\eq{nn}
\n^k \; \mbox{for $k\geq0$ \quad and}\quad (\n+\one)^k \; \mbox{for $k<0$}
\en
of the boson number operator $\n$ from a probabilistic point of view. We consider the ground state $\gr$
in Nelson's scalar quantum field model, and propose a representation of the ground state expectations
$(\gr, \n^k\gr)$, $k \geq 0$, and $(\gr, (\n+\one)^k\gr)$, $k<0$, in terms of a Gibbs measure $\cn$ on path
space developed in \cite{bhlms} and a random-time Poisson process $\proo N$. Note that estimating
$(\gr, \n\gr)$ is the cornerstone in the study the ground state properties, containing fundamental information
on the behaviour of a particle interacting with the field. By our method presented below we are able to obtain
similar estimates for both arbitrary integer and non-integer powers of the ground state expectations of the
number operator.
However, we note that while our initial purpose was to derive estimates, in fact we obtain equalities between
functionals of a Fock space operator and expectations of the joint Gibbs-Poisson measure (see Theorem 2.3),
and we believe that this method also has an interest in more general mathematics than just the confines of the
specific model.

%Apart from its relevance in mathematical physics we believe that it is a method which has
%an interest in more general mathematics as well, beyond the confines of the specific model.

Let $g$ be the coupling constant in the sum of the free and the interaction terms in the Nelson model. It is
known that
\eq{l2}
(\gr, \n\gr)=O( \g)
\en
for large $g$, see for details \kak{5} below. This can be derived from the so called pull-through formula and
extended to $(\gr, \n^n\gr)$ for $n\in\NN $ by a simple combinatorial argument. We extend this to all powers
$k\in\RR$, and show that
\eq{j0}
(\gr, \n^k\gr)= O(g^{2k}),\quad k\geq1
\en
in the strong coupling limit.

A basic idea is to take $k=m+\frac{\alpha}{2}\in\RR$ with  $m\in\ZZ$ and $0\leq \alpha <2$, and write
\eq{j1}
\n^k=\n^m\n^\frac{\alpha}{2}=\n^m \int_0^\infty (1-e^{-\beta \n})\la(d\beta)
\en
with a suitable Borel (in fact, L\'evy) measure $\la$. We have, furthermore, that
\eq{j2}
\n^m=(-1)^m\frac{d^m}{d\beta^m}e^{-\beta\n}\lceil_{\beta=0},\quad m>0,
\en
and
\eq{j3}
\n^m=\int_{[0,\infty)^m} e^{-(\beta_1+\cdots+\beta_m) \n}\prod_{j=1}^md\beta_j,\quad m<0.
\en
Combining \kak{j1}, \kak{j2} and \kak{j3} with the path integral representation of $(\gr, e^{-\beta\n}\gr)$
given in terms of a Gibbs measure $\cn$, we derive the probabilistic representation of $(\gr, \n^k\gr)$
for all $k\in\RR$.
Using this expression we obtain the asymptotic behaviour
$$
C_1\leq \lim_{g\to\infty} \frac{(\gr, \n^k\gr)}{g^{2k}} \leq C_2,\quad k\geq1,
$$
with some constants $C_1,C_2>0$, as a corollary.

In the remainder of this section we recall the definition and some basic facts on the Nelson model. In
Section 2 we state the main theorem and its proof. In Section 3 we discuss an application to the polaron
model, which is related to the Nelson model in a specific sense.

\subsection{Nelson model}
For background material we refer to \cite{LHB11}. We consider the Nelson model of an
electrically charged spinless quantum particle coupled to a scalar boson field. The
particle in assumed to be under an external potential $V$, and its Hamilton operator
is described by the Schr\"odinger operator
\begin{equation}
\label{schr11}
\hp = -\frac{1}{2} \Delta + V.
\end{equation}
We choose $V$ such that $\hp$ has a unique strictly positive ground state $\grp \in L^2(\BR)$.

To describe the scalar quantum field consider the boson Fock space $\fff=\mathbb C \oplus
\oplus_{n=1}^\infty \otimes _{\rm sym}^n \LR$ over $\LR$, where the subscript indicates
symmetrized tensor product.
%We denote the $n$-particle subspace, called $n$th Fock sector, by $\fff^{(n)}=\otimes_{\rm sym}^n \LR$, and define $\fff^{(0)} = \mathbb C$.
The free field Hamilton operator is given %(in a formal notation)
by \eq{a2}
\hf=\int_\BR\omega(k)\add(k)a(k) dk,
\en
where $a(k)$ and $\add(k)$ are the boson annihilation and creation operators, respectively,
satisfying $[a(k),\add(k')]=\delta(k-k')$, and
\eq{massive}
\omega(k)=\sqrt{|k|^2+\nu^2},\quad \nu\geq 0
\en
is the dispersion relation, with boson mass $\nu$. Finally, the Hamilton operator of the
interaction term between particle and field is
\eq{1}
\phi(x)=\frac{1}{\sqrt2}\int_\BR \lk\frac{\vp(k)}{\sqrt{\omega(k)}}e^{-ik \cdot x} \otimes\add(k) +
\frac{\vp(-k)}{\sqrt{\omega(k)}}e^{+ik\cdot x}\otimes a(k)\rk dk,
\en
where $\vp$ is the Fourier transform of a given function $\varphi$ describing the charge distribution
of the particle. We assume throughout this paper that $\varphi(x)\geq 0$.
%\eq{a}
%\varphi(x)\geq 0.
%\en
The charge distribution $\varphi$ regularizes the particle from a point charge and imposes an ultraviolet
cutoff making the interaction well defined. The Nelson Hamiltonian
\eq{3}
\hn=\hp\otimes \one+\one\otimes \hf+g\phi
\en
is defined on the Hilbert space $\hhh=\LR\otimes\fff$, with coupling constant $g\in\RR$.

Under the assumption that $\sqrt\om\vp\in\LR$ and $\vp/\sqrt\om\in\LR$ it can be proven that $\hn$ is
a self-adjoint operator on $D(\hp)\cap D(\hf)$. Denote
\eq{4}
\II=\half \int_\BR\frac{|\vp(k)|^2}{\omega(k)^3}dk < \infty.
\en
It is well known that $\hn$ has a unique strictly positive ground state $\gr \in L^2(\BR) \otimes \fff$
for every $g \in \RR$ whenever $\nu>0$ \cite{ara01, bfs, G,hhs, sasa, Sp}, and it has no ground state
in this space if $\nu=0$ \cite{dg1, hi06,lms} unless \kak{4} holds, imposing an infrared condition.
Throughout this paper we assume that $\nu>0$.

Let
\eq{nnn}
\n = \int_\BR \add(k) a(k) dk
\en
be the boson number operator acting in $\fff$. The ground state expectation of $\n$ is very useful in
studying the spectrum of the Nelson model and it is given by
\eq{5}
(\gr, \n\gr)=\frac{g^2}{2} \int_\BR \left \|(\hn-E+\omega(k))\f\frac{\vp(k)}{\sqrt{\omega(k)}}
e^{-ik\cdot x} \gr\right\|_\hhh^2 dk,
\en
where $E$ is the infimum of the spectrum of $H$. The right hand side is bounded by \kak{4}. This is
derived from the so called pull-through formula
$$
a(k)\gr=\frac{g}{\sqrt2}(\hn-E+\om(k))\f e^{-ik\cdot x} \frac{\vp(k)}{\sqrt{\omega(k)}}\gr.
$$
In this paper our main objective is to estimate $(\gr, \n^k \gr)$, for all $k\in\RR$.

\section{Main result and proofs}
\subsection{Main theorem}
Consider the class of completely monotone functions
$$
\ms B=\left\{ f\in C^\infty(\RR^+)\, \left|\, f(x) \geq 0, \,
(-1)^n \frac{d^nf}{dx^n}(x) \leq 0, \, \forall n\in\NN \right.\right\}.
$$
An element of $\ms B$ is called a Bernstein function. Bernstein functions are positive, increasing
and concave. An example is $\bern(u)=cu^{\TT  }$, $c\geq 0$, $0<\alpha\leq 2$.

Let $\ms L$ be the set of Borel measures $\lambda$ on $\RR\setminus\{0\}$ such that
\begin{enumerate}
\item[(1)]
$\lambda((-\infty,0))=0$
\item[(2)]
$\d \int_{\RR\setminus\{0\}} (y\wedge1)\lambda (dy)<\infty$.
\end{enumerate}
For every Bernstein function $\bern\in \BERN $ there exists a unique pair $(b,\lambda )\in
\RR^+ \times \ms L$ such that
\eq{6}
\bern(u)=bu+\int_0^\infty (1-e^{-uy})\lambda (dy).
\en
Conversely, the right hand side of \kak{6} is an element of $\BERN $ for each pair $(b,\lambda )
\in \RR^+\times \ms L$.

Instead of considering just $(\gr, \n^k\gr)$ we study the more general functionals
\eq{16}
(\gr, \n^m\Psi(\n)\gr),\quad m\in \NN \cup\{0\}, \quad \Psi\in\BERN_0
\en
and
\eq{16a}
(\gr, (\n+\one)^{-m}\Psi(\n+\one)\gr),\quad m\in \NN , \quad \Psi\in\BERN_0.
\en
Note that \kak{16} and \kak{16a} include the cases $(\gr, \n^k\gr)$, $k\geq0$, and
$(\gr, (\n+\one)^k\gr)$, $k<0$, obtained for $\Psi(u)=u^{\TT  }$, $0\leq \alpha <2$.
Since $u^m\Psi(u)=b u^{m+1}+u^m \int_0^\infty (1-e^{-uy})\la(dy)$, we can disregard the
constant (drift) part and it suffices to work with the subclass
$$
\BERN_0 =\lkk f\in \ms B \left|b=0 \right.\rkk
$$
and derive a formula for $\Psi\in\BERN_0$.

Define
\eq{8}
\rho(\beta)=(\gr, e^{-\beta \n}\gr),\quad \beta\in\CC.
\en
In \cite[Chapter 6]{LHB11} it has been established that $\rho(\beta)$ is analytic and it
is represented in  terms of a Gibbs measure $\cn$ on the path space $\ms X =C(\RR;\BR)$ of
continuous functions. We give an outline of this relationship for the convenience of the
reader.

\begin{proposition}
\label{10}
There exist a probability measure $\cn$ and a diffusion process $\pro X$ on $(\ms X, \ms G,\cn)$
such that
\eq{nn1}
\rho(\beta)=\Ebb_\cn\left[e^{-\g (1-e^{-\beta})W}\right],\quad\beta\in\CC,
\en
where
$$
W = W(\omega) = \int_{-\infty}^0ds\int_0^\infty \ms W(X_t(\omega)-X_s(\omega),t-s) dt,
\quad \omega\in\ms X ,
$$
and $\ms W(x,t)$ is a pair interaction potential given by
$$
\ms W(x,t)=\half \int_\BR \frac{|\vp(k)|^2}{\omega(k)} e^{-ik \cdot x}e^{-|t|\omega(k)} dk.
$$
\end{proposition}
\proof
For a detailed proof we refer to \cite[Chapter 6]{LHB11}. Recall that $\grp > 0$ is the ground state of
$\hp$, and define the unitary operator $U:L^2(\BR,\grp^2 dx)\to\LR$ by $f\to\grp f$. Then there exists
a diffusion process $\pro X$ on $(\ms X,\ms G, \cn_0)$, whose generator is the self-adjoint operator
$U\f \hp U$. The Nelson Hamiltonian $\hn$ is transformed by $U\f\otimes \one$ to the self-adjoint operator
$$U\f \hp U\otimes\one+\one\otimes\hf+g\phi$$ on $L^2(\BR;\grp^2 dx)\otimes \fff$. We denote it by the same
$\hn$. Since the ground state of $\hn$ is strictly positive, $(\gr, \one\otimes\OF )>0$, where $\OF =
(1, 0, 0, ...)$ denotes the Fock vacuum. Thus we have in the $T\to\infty$ limit
$$
\grt= (\|e^{-T\hn}(\one\otimes\OF )\|)\f e^{-T\hn}(\one\otimes\OF ) \to\gr
$$
in strong sense, and we obtain that $(\gr,e^{-\beta \n}\gr)=\lim_{T\to\infty}(\grt,e^{-\beta \n}\grt)$ for
$\beta>0$. On the other hand, the path integral representation
\begin{eqnarray}
&&\hspace{-2cm}
(\grt,e^{-\beta \n}\grt)  \nonumber \\ &&
\hspace{-2cm}
=\frac{1}{Z_T}\int e^{-\g\int_{-T}^0\int_0^T W(X_t(\omega)-X_s(\omega),t-s)
(1-e^{-\beta})}
e^{\frac{\g}{2}  \int_{-T}^T\int_{-T}^T W(X_t(\omega)-X_s(\omega),t-s)}d\cn_0(\omega)
\label{int}
\end{eqnarray}
holds, where
$$
Z_T = \int e^{\frac{\g}{2}  \int_{-T}^T\int_{-T}^T W(X_t(\omega)-X_s(\omega),t-s)}d\cn_0(\omega)
$$
is the normalizing factor. We can prove that the family of probability measures
$$
d\cn_T=\frac{1}{Z_T}e^{\frac{\g}{2} \int_{-T}^T \int_{-T}^T W(X_t-X_s,t-s)}d\cn_0, \quad T\geq 0,
$$
is tight, and the integrand $e^{-\g \int_{-T}^0\int_0^T W(X_t-X_s, t-s)(1-e^{-\beta})}$ in \kak{int}
is uniformly convergent with respect to the paths $\omega$ as $T\to\infty$. Hence there exits a
subsequence $\seq {T'}$ such that $\cn_{T'_n}$ weakly converges to a probability measure $\cn$, and
\kak{nn1} is satisfied for $\beta>0$. This can be further extended to all $\beta\in\CC$ by analytic
continuation. $\cn$ is a Gibbs measure on path space for potential $V$ contained in $\hp$ and the
pair interaction $\ms W$ above.
\qed

\bl{fundamental}
For every path $\omega \in \ms X$ we have $0\leq W(\omega) \leq \II$.
\enl
\proof
The upper bound is straightforward. By Fourier transformation
\begin{eqnarray}
\ms W(x-y, t)
&=&
(\varphi(\cdot-x), \om(-i\nabla)\f e^{-|t|\om(-i\nabla)}\varphi(\cdot-y)) \nonumber \\
&=&
\int_{|t|}^\infty (\varphi(\cdot-x),  e^{-s\om(-i\nabla)}\varphi(\cdot-y))ds,
\end{eqnarray}
for every $x, y \in \BR$ and $t\in\RR$, and where $\omega$ in the equality above is the dispersion relation.
Since $e^{-s\om(-i\nabla)}$ is positivity improving and $\varphi\geq0$ by assumption, we conclude that
$\ms W(x-y,t)\geq0$.
\qed
Let $\proo N$ be a Poisson process with intensity $1$ on a probability space $(\Omega , {\cal F}, P$).
Recall that
\begin{equation}
\label{poi}
\Ebb_P[e^{-uN_t}] = e^{t(e^{-u}-1)}, \quad t \geq 0,
\end{equation}
and write
$$
\Ebb=\Ebb_{\cn}\Ebb_P.
$$
The main result of this paper is as follows.
\bt{main1}
Let $m\in \mathbb Z$ and $\Psi\in \BERN_0$. Then
\begin{eqnarray}
\label{hh1}
\d m\geq 1 \; & \Longrightarrow & \; \d (\gr, \n^m\Psi(\n)\gr) =
\sum_{r=1}^m S(m,r) \Ebb\left[\lk\g W\rk^r \Psi(N_{\g W}+r)\right]\\
\label{hh2}
\d m = 0 \; & \Longrightarrow &
\d \; (\gr, \Psi(\n)\gr) = \Ebb \left[\Psi(N_{\g W})\right]\\
\d m \leq -1 \; & \Longrightarrow & \; \d (\gr, (\n+\one)^{m}\Psi(\n+\one)\gr)
= \d\Ebb\left[(N_{\g W}+1)^{m} \Psi(N_{\g W}+1) \right] \nonumber \\
\label{hh3}
\end{eqnarray}
where $S(m,r)$ are the Stirling numbers of the second kind, and $N_{\g W}$ is $N_t$ evaluated at
the random time $t=\g W$.
\et
In order to avoid singularities, in Theorem \ref{main1} we replaced $\n$ by $\n+\one$ for the case
$m<0$, however, any positive multiple of $\one$ can be used. In particular, we have the following
formulae.
\bc{11}
Let $m\in \ZZ $ and $0\leq \alpha<2$. Then
\begin{eqnarray}
\label{hh4}
m\geq 1 \; & \Longrightarrow & \; \d (\gr, \n^{m+\frac{\alpha}{2}}\gr)  =
\sum_{r=1}^m S(m,r) \Ebb\left[\lk\g W\rk^r (N_{\g W}+r)^{\TT}\right]\\
\label{hh5}
m = 0 \; & \Longrightarrow & \; \d (\gr, \n^{\TT  }\gr)=
\Ebb \left[(N_{\g W})^{\TT  }\right]\\
\label{hh6}
m \leq -1 \; & \Longrightarrow & \; \d(\gr, (\n+\one)^{m+\frac{\alpha}{2}}\gr) =
\d\Ebb\left[(N_{\g W}+1)^{m+\TT  } \right].
\end{eqnarray}
\ec

Moreover, for purely natural powers of the number operator we obtain the following expression.
\bc{111}
For $\alpha = 0$ and $m \in \mathbb{N}$ we have
\begin{equation}
(\gr, \n^{m}\gr) = \Ebb_\cn \Ebb_{P^\omega} [X^m],
\end{equation}
where $X$ is a Poisson random variable on a probability space with random intensity $\g W(\omega)$,
independent of $W(\omega)$, and whose distribution we denote by $P^\omega$.
\ec
\begin{proof}
From \kak{hh4} we have
$$
(\gr, \n^{m}\gr)  = \sum_{r=1}^m S(m,r) \Ebb\left[\lk\g W\rk^r\right] =
\Ebb_\cn\left[\sum_{r=1}^m S(m,r)\lk\g W\rk^r\right].
$$
Let $X$ be a Poisson random variable with intensity $\mu$ and probability distribution $P^\mu$ on a
suitable probability space, and recall the formula
$$
\Ebb_{P^\mu} [X^m] = \sum_{r=1}^m S(m,r) \mu^r, \quad  m \in \mathbb{N}.
$$
Then the claim follows for $\mu = \g W(\omega)$ and $P^\mu = P^\omega$.
\qed
\end{proof}

\subsection{Proof of Theorem \ref{main1}}
We first prove the case $m\geq0$.

{\it Proof of \kak{hh1} and \kak{hh2}}:
It suffices to consider only $m\geq1$. There exists a L\'evy measure $\lambda$ such that
\eq{j10}
\Psi(u)=\int_0^\infty (1-e^{-uy})\la(dy),
\en
and
\eq{j11}
\n^m=(-1)^m\frac{d^m}{d\beta^m}e^{-\beta\n}\lceil_{\beta=0}.
\en
By Proposition \ref{10} and the combination of \kak{j10} and \kak{j11}
we have
$$
(\gr, \n^m\Psi(\n)\gr)
=(-1)^m\int_0^\infty(\partial^m \rho(0)-\partial^m \rho(\beta))\la(d\beta).
$$
A direct calculation gives
$$
\partial^m \rho(\beta) = (-1)^m \sum_{r=1}^m  a_r(m) e^{-r\beta}
\Ebb_{\cn}\left[\lk\g W\rk^r e^{-\g W(1-e^{-\beta})} \right],
$$
where
\eq{j5}
\d a_r(m)=\frac{(-1)^r}{r!}
\sum_{s=1}^r(-1)^s\lk\!\!
\begin{array}{l}
r\\ s
\end{array}
\!\!\rk s^m.
\en
It is well-known \cite{Lov07} that the sum in \kak{j5} can be expressed in terms of the
Stirling numbers of the second kind $S(m,r)$, and thus
$$
a_r(m) =
 \left\{
 \begin{array}{ll}
  0 & {\rm if} \;\; 0 \leq m < r \medskip\\
  S(m,r) & {\rm if} \;\; m \geq r.
\end{array} \right.
$$
Note that $S(m,m)=1$, for all $m \in \mathbb{N}$. This yields
\begin{eqnarray*}
(\gr, \n^m\Psi(\n)\gr) = \Ebb_{\cn}\left[\int_0^\infty \sum_{r=1}^m S(m,r) \lk\g W\rk^r (1-e^{-r\beta}
e^{-\g W(1-e^{-\beta})}) \la(d\beta) \right].
\end{eqnarray*}
Using (\ref{poi}) gives
$$
1-e^{-r\beta}e^{-\g W(1-e^{-\beta})} =\Ebb_P \left[1-e^{-\beta(N_{\g W}+r)}\right].
$$
Hence
$$
(\gr, \n^m\Psi(\n)\gr) = \Ebb_{\cn}\left[\sum_{r=1}^m S(m,r) \lk\g W\rk^r
\Ebb_P\left[\int_0^\infty (1-e^{-\beta(N_{\g W}+r)}) \la(d\beta)\right] \right],
$$
and the theorem follows.
\qed

Next we consider the case $m<0$. The strategy is similar to the case of positive powers.

{\it Proof of \kak{hh3}}:
By a combination of the Laplace transform
\eq{la1}
\int_{[0,\infty)^m} \prod_{j=1}^m d\beta_j e^{-\sum_{j=1}^m \beta_j (\n+\one)} = (\n+\one)^{-m}
\en
and
\eq{la2}
\Psi(\n+\one)=\int_0^\infty (1-e^{-\beta(\n+\one)}) \lambda(d\beta),
\en
we obtain
\begin{eqnarray*}
\label{hi1}
\lefteqn{
(\gr, (\n+\one)^{-m}\Psi(\n+\one)\gr) } \\ &&
= \int_{[0,\infty)^m} \prod_{j=1}^m d\beta_j \int_0^\infty \lambda(d\beta)
(\gr, (e^{-\sum_{j=1}^m \beta_j (\n+\one)}- e^{-(\sum_{j=1}^m \beta_j+\beta)(\n+\one)})\gr).
\end{eqnarray*}
By \kak{hi1} it follows that
\begin{eqnarray*}
\lefteqn{
(\gr, (\n+\one)^{-m}\Psi(\n+\one)\gr) } \\
&=&
\int_{[0,\infty)^m} \prod_{j=1}^m d\beta_j \int_0^\infty \lambda(d\beta) e^{-\sum_{j=1}^m \beta_j}
\Ebb_{\cn}\left[ e^{-\g W(1-e^{-\sum_{j=1}^m \beta_j})} -
e^{-\g W(1-e^{-\sum_{j=1}^m \beta_j+\beta})} e^{-\beta} \right].
\end{eqnarray*}
In terms of the Poisson process $\proo N$ we can rewrite as
\begin{eqnarray*}
&=&
\int_{[0,\infty)^m} \prod_{j=1}^m d\beta_j \int_0^\infty\lambda(d\beta)
\Ebb_{\cn}\Ebb_P\left[e^{-\sum_{j=1}^m \beta_j(N_{\g W}+1)} -
e^{-(\sum_{j=1}^m \beta_j+\beta) (N_{\g W}+1)} \right]\\
&=&
\int_{[0,\infty)^m} \prod_{j=1}^m d\beta_j \int_0^\infty\lambda(d\beta)
\Ebb_{\cn} \Ebb_P \left[e^{- \sum_{j=1}^m \beta_j(N_{\g W}+1)}
(1- e^{- \beta (N_{\g W}+1)}) \right].
\end{eqnarray*}
Integrating with respect to $\la(d\beta)$ and then $\prod_j d\beta_j$, we finally obtain
\begin{eqnarray*}
&=&
\int_{[0,\infty)^m} \prod_{j=1}^m d\beta_j
\Ebb_{\cn} \Ebb_P\left[e^{-\sum_{j=1}^m \beta_j(N_{\g W}+1)}\Psi(N_{\g W}+1)
\right]\\
&=&
\Ebb\left[(N_{\g W}+1)^{-m}\Psi(N_{\g W}+1)\right].
\end{eqnarray*}
Hence the theorem follows.
\qed

\begin{remark}
\rm{
Since the coefficient $a_r(m)=1$ for $1=r=m$,  we note that formula \kak{5} can actually also
be obtained from $(\gr, \n\gr)=\Ebb[\g W]$. We have
\eq{ko}
\Ebb[\g W]=\frac{\g}{2} \int_\BR dk \frac{|\vp(k)|^2}{\omega(k)} \int_{-\infty}^0dt\int_0^\infty
e^{-|t-s|\omega(k)} \Ebb[e^{-ik \cdot (X_s-X_t)}] ds.
\en
Furthermore,
$$
\Ebb[e^{-ik \cdot (X_s-X_t)}] = \Ebb_{\cn}[e^{-ik \cdot (X_s-X_t)}]=
(e^{-ik \cdot x}\gr, e^{-|t-s|(\hn-E)}e^{-ik\cdot x}\gr).
$$
Inserting this into \kak{ko} gives \kak{5}.
}
\end{remark}

\subsection{Lower and upper bounds}
The above theorem allows to derive the asymptotic behaviour of the ground state expectation of non-integer
powers of the boson number operator in the strong coupling limit.
\bc{13}
Let $k=m+\frac{\alpha}{2}\geq 1$, $m\in\NN $ and $0\leq \alpha<2$. Suppose that there exists $a>0$ not
depending on $g$, and $\Ebb[W]\geq\II-a>0$. Then
\eq{l1-1}
(\II-a)^k \leq \lim_{g\to\infty} \frac{(\gr, \n^k\gr)}{g^{2k}} \leq \II^k.
\en
\ec
\proof
By Jensen's inequality we have
$$
\Ebb_P \left[(N_{\g W}+r)^{\TT  }\right] \leq (\Ebb_P [N_{\g W}+r])^{\TT  } = (\g \II+r)^{\TT  }.
$$
In particular, it follows that
$$
(\gr, \n^{m+\frac{\alpha}{2}}\gr) \leq \sum_{r=1}^m S(m,r) (\g \II+r)^{\TT  } (\g \II)^r,
$$
implying $\lim_{g\to\infty}(\gr, \n^{m+\frac{\alpha}{2}}\gr)/g^{2m+\alpha} \leq S(m,m)\II^k = \II^k$.
This gives the upper bound. Again, from Jensen's inequality it follows that
$$
(\gr, \n^k\gr) \geq  (\gr, \n\gr)^k,
$$
Since $(\gr, \n\gr)=\g \Ebb_{\cn}[W]$, the lower bound follows.

\begin{remark}
\rm{
Since any Bernstein function $\Psi$ is increasing and concave, by the Jensen inequality and Lemma
\ref{fundamental} we obtain more generally that
\begin{eqnarray*}
(\gr, \Psi(\n)\gr)
&=&
\int \Ebb_P[\Psi(N_{\g W})] d\cn \leq \int \Psi(\Ebb_P[N_{\g W}]) d\cn \\
&=&
\int \Psi(\g W)  d\cn \leq \Psi(\g W_\infty).
\end{eqnarray*}
}
\end{remark}

\begin{remark}
{\rm
The assumption stated in Corollary \ref{13} is fairly standard. Let $V=V_1-V_2$, where $V_2\geq0$, $V_2\in
L^p(\BR)$, $p=1$ for $d=1$, $p>d/2$ for $d\geq 2$, and $V_1\in L_{\rm {loc}}^1(\BR)$, $\inf_x V_1(x)>
-\infty$. Denote $\Sigma=\liminf_x V(x)$, and assume
\bi
\item[(1)]
$V_1(x)\to\infty$ as $|x|\to\infty$,
\item[(2)]
$\Sigma> \max\{E, \inf_x V_1(x)\}$.
\ei
Under either (1) or (2) the ground state $\|\gr(x)\|$ decays exponentially, moreover
$$
\||x|\gr(x)\|_{\fff}\leq C\|\gr\|_{\hhh}
$$
with a suitable constant $C > 0$ independent of $g$ (for details see \cite[Chapter 6]{LHB11}). Let
$\sup_x \||x|\gr(x)\|_{\fff}<C\|\gr\|<\infty$. Then it is also shown in the same reference that
\eq{hh7}
\Ebb[W]\geq \half\int_\BR \frac{\vp(k)^2}{\omega(k)^3} (1-C|k|^2) dk.
\en
(Remember that $\|\gr\|=1$.) If the field mass $\nu>0$ is sufficiently small, $d\leq 3$ and $\vp$ is
continuous in a neighborhood of $k=0$, then the right hand side of \kak{hh7} is strictly positive.
}
\end{remark}

\section{Applications}
\subsection{Polaron model}
A related model is the polaron model obtained for dispersion relation $\omega(k) = 1$. The same result
as in the Nelson model can be obtained by using a similar argument.

The Hamiltonian of the polaron is defined by
\eq{p1}
\Hp=\hp\otimes\one+\one\otimes\n+g\phi,
\en
where
\eq{p2}
\phi(x)=\frac{1}{\sqrt2}\int_\BR \lk\frac{\vp(k)}{|k|} e^{-ik\cdot x} \otimes\add(k) +
\frac{\vp(-k)}{|k|}e^{+ik\cdot x}\otimes a(k)\rk dk.
\en
If $\hat\varphi/|k|\in\LR$, then $\Hp$ is self-adjoint on $D(\hp)\cap D(\n)$. We assume $\varphi(x)\geq0$,
and choose $V$ such that $\hp$ has a strictly positive ground state $\grp$. Also, we assume that
\eq{p3}
\III=\half \int_\BR\frac{|\vp(k)|^2}{|k|^2}dk<\infty.
\en
Note that whenever $d\geq 3$, there is no infrared divergence. Then under \kak{p3} $\Hp$ has a unique
ground state $\gr > 0$ for every $g \in \RR$. In a similar way to Proposition \ref{10} we obtain that there
exist a probability measure $\cn$ and a diffusion process $\pro X$ on the probability space $(\ms X, \ms G,\cn)$
such that
\eq{nn10}
(\gr, e^{-\beta \n}\gr)=\Ebb_\cn\left[e^{-\g (1-e^{-\beta})W}\right],\quad\beta\in\CC,
\en
where
$$
W_{\rm {pol}}(\omega) = \int_{-\infty}^0ds\int_0^\infty \ms W_{\rm {pol}}(X_t(\omega)-X_s(\omega),t-s) dt
$$
with pair interaction potential
$$
\ms W_{\rm{pol}}(x,t)=\half e^{-|t|} \int_\BR \frac{|\vp(k)|^2}{|k|^2}e^{-ik \cdot x}dk.
$$
In particular, for $\vp(k)=(2\pi)^{3/2}$ and $d=3$ we have
\eq{pp4}
W_\p(\omega)= \frac{1}{8\pi}\int_{-\infty}^0ds\int_0^\infty\frac{e^{-|t-s|}}{|X_t(\omega)-X_s(\omega)|}dt.
\en
\bl{pp5}
$0\leq W_\p (\omega)\leq \III$.
\enl
\proof
The upper bound is again straightforward. Fourier transformation gives
$$
\ms W_\p(x-y,t)=(\varphi(\cdot-x), \Delta\f\varphi(\cdot-y))e^{-|t|}.
$$
Since $\Delta\f=\int_0^\infty e^{\beta\Delta}d\beta$ and $e^{\beta\Delta}$ is positivity preserving for
$\beta\geq0$, the lemma follows as $\varphi$ is positive by assumption.
\qed
Hence we have a similar result to Theorem \ref{main1}.
\bc{main111}
Let $m\in \mathbb Z$ and $\Psi\in \BERN_0$. Then
\begin{eqnarray*}
\label{hhh1}
m\geq 1 \; & \Longrightarrow & \; (\gr, \n^m\Psi(\n)\gr) = \sum_{r=1}^m S(m,r)
\Ebb\left[\lk\g W_\p\rk^r \Psi(N_{\g W_\p}+r)\right]\\
\label{hhh2}
m = 0 \; & \Longrightarrow & \; (\gr, \Psi(\n)\gr) = \Ebb \left[\Psi(N_{\g W_\p})\right]\\
m \leq -1 \; & \Longrightarrow & \; (\gr, (\n+\one)^{m}\Psi(\n+\one)\gr)
= \d\Ebb\left[(N_{\g W_\p}+1)^{m} \Psi(N_{\g W_\p}+1) \right].\nonumber \\
\label{hhh3}
\end{eqnarray*}
\ec

\subsection{Nelson model with zero total momentum}
Finally we consider the Nelson  model $\hn$ with zero external potential $V=0$. In this case $\hn$ is
translation invariant and can be decomposed in terms of the spectrum of the total momentum
$$
-i\nabla\otimes\one+\one\otimes P_{\rm f},
$$
where $P_{\rm f}=\int_\BR k\add(k) a(k) dk$ is the field momentum operator. Thus the decomposition
\eq{t1}
H=\int^\oplus_\BR \hn(P) dP,\quad \hhh=\int^\oplus_\BR\hhh(P) dP
\en
holds, where
\eq{t2}
H(P)=\half (P-P_{\rm f})^2+g\phi(0)+\hf
\en
and
%\eq{t3}
$\hhh(P)=\fff$.
%\en
As in the case of the Nelson model we assume that $\omega(k)=\sqrt{|k|^2+\nu^2}$ with $\nu >0$,
and $\varphi(x)\geq0$. It follows that
\eq{t13}
(F, e^{-t\hn(P)}G)=
\Ebb_{{\cal W}^0}\left[(F_0, e^{-g\phi_E(\int_0^t \tilde \varphi(\cdot-B_s)ds)}
e^{i(P-P_{\rm f})\cdot B_t}G_t)_{\fff}\right].
\en
Here $(B_t)_{t\in\RR}$ is two-sided Brownian motion on $\RR$, and ${\cal W}^0$ is Wiener measure
starting from zero on $(\ms X,\mathcal F)$, where $\ms X=C(\RR;\BR)$. Furthermore, $\tilde \varphi$ 
is the Fourier transform of $\vp/\sqrt \omega$, $\phi_E$ is the Euclidean scalar field, and $F_0$ and   
$G_t$ are the projections at time zero and time $t$ of the Euclidean scalar field of $F$ and $G$, 
respectively. For further details we refer to \cite{hir07}. Since $e^{-iP_{\rm f}\cdot B_T}\Omega_\fff=
\Omega_\fff$ for all $T\in\RR$, taking $F=G=\Omega_\fff$ we obtain that
\eq{t1333}
(\Omega_\fff, e^{-T\hn(P)}\Omega_\fff)=\Ebb_{{\cal W}^0}\left[e^{iP\cdot B_T}
e^{\g \int_0^Tds\int_0^T dt \ms W(B_t-B_s, t-s)}\right].
\en

Now put $P=0$. Then the phase $e^{iP\cdot B_T}$ in \kak{t13} becomes 1, and we see that $e^{-t\hn(0)}$ 
is positivity improving since $e^{-iP\cdot P_{\rm f}}$ is a shift operator. Thus the ground state of $\hn(0)$ 
is unique and strictly positive; we denote it by $\gr(0)$. We have then that
$$
(\gr(0),e^{-\beta \n}\gr(0))=\lim_{T\to\infty}(\grt(0), e^{-\beta \n}\grt(0))
$$ 
for $\beta>0$, where $\grt(0)=e^{-T\hn(0)}\Omega_\fff/\|e^{-TH(0)}\Omega_\fff\|$. Its path integral 
representation is
\eq{t4}
(\grt(0), e^{-\beta \n}\grt(0))=\Ebb_{\mu_T}\left[e^{-\g(1-e^{-\beta})W_T}\right],
\en
where
\eq{t11}
W_T = W_T(\omega) = \int_{-T}^0 dt\int_0^T \ms W(B_t(\omega)-B_s(\omega),t-s) ds
\en
and
\eq{t6}
d\mu_T=\frac{1}{Z_T}\exp\lk\frac{\g}{2}\int_{-T}^T dt \int_{-T}^T \ms W(B_t-B_s, t-s) ds \rk
d{\cal W}^0.
\en
Denote $\mathcal F_{[-T,T]}=\sigma(B_s,-T\leq s\leq T)$ for $T >0$.
\begin{proposition}
\label
{Gibbs}
There exists a probability measure $\mu$ on $(\ms X, \mathcal F)$ such that for every $\fff_{[-T,T]}$-measurable
function $f$, $T>0$,
\eq{t7}
\lim_{T\to\infty}\Ebb_{\mu^T}[f]=\Ebb_\mu[f].
\en\end{proposition}
\proof See \cite{bs05} and \cite[Theorem 4.36]{LHB11}.
\qed

\bt{hiroshima}
Let $\beta\in\CC$. Then
\eq{t8}
(\gr(0), e^{-\beta \n}\gr(0))=\Ebb_\mu\left[e^{-\g(1-e^{-\beta})W}\right].
\en
\et
\proof
Suppose $\beta>0$ and write $S_T=-\g(1-e^{-\beta})\int_{-T}^0 dt \int_0^T \ms W(B_t-B_s,t-s) ds$. Note again 
that $\lim_{T\to\infty} S_T= S_\infty$, uniformly in paths. For every $\varepsilon>$ there exists $T_0>0$ such 
that $|e^{S_T}-e^{S_\infty}|<\varepsilon$ for all $T>T_0$, uniformly in paths. Fix $\varepsilon$ and $T_0$, and 
choose $t>T>T_0$. By telescoping
\begin{eqnarray*}
\lefteqn{
\Ebb_{\mu}[e^{S_\infty}]-\Ebb_{\mu_T}[e^{S_T}] } \\ &&
=(\Ebb_{\mu}[e^{S_\infty}]-\Ebb_{\mu}[e^{S_t}])+ (\Ebb_{\mu}[e^{S_t}]-\Ebb_{\mu_T}[e^{S_t}]) +
(\Ebb_{\mu_T}[e^{S_t}]-\Ebb_{\mu_T}[e^{S_T}]),
\end{eqnarray*}
we obtain
\eq{t9}
|\Ebb_{\mu}[e^{S_\infty}]-\Ebb_{\mu_T}[e^{S_T}]|\leq \varepsilon + |\Ebb_{\mu}[e^{S_t}]-\Ebb_{\mu_T}[e^{S_t}]|
+2\varepsilon.
\en
By Proposition \ref{Gibbs}, the second term at the right hand side of \kak{t9} converges to zero as $t\to\infty$.
Thus the theorem follows for $\beta>0$. By analytic continuation the theorem follows for all $\beta\in \CC$.
\qed
Write $\Ebb=\Ebb_{{\cal W}^0}\Ebb_P$. The next corollary is immediate.
\bc{main11}
Let $m\in \mathbb Z$ and $\Psi\in \BERN_0$. Then
\begin{eqnarray*}
\label{hhhh1}
m\geq 1 \; & \Longrightarrow & \; (\gr(0), \n^m\Psi(\n)\gr(0)) = \sum_{r=1}^m S(m,r)
\Ebb\left[\lk\g W\rk^r \Psi(N_{\g W}+r)\right]\\
\label{hhhh2}
m = 0 \; & \Longrightarrow & \; (\gr(0), \Psi(\n)\gr(0)) = \Ebb \left[\Psi(N_{\g W})\right]\\
m \leq -1 \; & \Longrightarrow & \; (\gr(0), (\n+\one)^{m}\Psi(\n+\one)\gr(0))
= \d\Ebb\left[(N_{\g W}+1)^{m} \Psi(N_{\g W}+1) \right].\nonumber \\
\label{hhhh3}
\end{eqnarray*}
\ec

\bigskip\bigskip
\noindent
\textbf{Acknowledgments}: FH acknowledges support of Grant-in-Aid for Science Research (B) 20340032
from JSPS and Grant-in-Aid for Challenging Exploratory Research 22654018 from JSPS.
We thank Itaru Sasaki for useful discussions and comments.

{\small
\begin{thebibliography}{99}
\bibitem[Ara01]{ara01}
A. Arai,
Ground state of the massless Nelson model without infrared cutoff in
a non-Fock representation, {\it Rev. Math. Phys.} {\bf 13} (2001),
1075--1094.



\bibitem[BFS98]{bfs}
V. Bach,  J.  Fr\"ohlich, and I. M. Sigal,
Quantum electrodynamics of confined non-relativistic particles,
{\it Adv. Math. } {\bf 137} (1998), 299--395.

\bibitem[BS05]{bs05}
V. Betz and H. Spohn, A central limit theorem for Gibbs measures relative to Brownian motion, 
{\it Probab. Theory Relate. Fields} {\bf 131} (2005), 459--478.

\bibitem[BHLMS02]{bhlms}
V.  Betz,  F. Hiroshima,  J.  L\H{o}rinczi, R.   Minlos  and H.  Spohn,
Ground state properties of the Nelson Hamiltonian: a Gibbs measure-based approach,
{\it Rev.  Math.  Phys.}  {\bf  14}  (2002),  173-198.

\bibitem[DG04]{dg1}
J. Derezi\'nski and C. G\'erard,
Scattering theory of infrared divergent Pauli-Fierz Hamiltonians,
{\it Ann. Henri Poincar\'e}, {\bf 5} (2004), 523-578.

\bibitem[Ger00]{G}
C. G{\'e}rard,
On the existence of ground states for massless Pauli-Fierz Hamiltonians,
{\it Ann. Henri Poincar\'e} {\bf 1} (2000), 443--459.


\bibitem[Hir06]{hi06}
M. Hirokawa,
Infrared catastrophe for Nelson's model: non-existence of ground state and soft-boson
divergence,
{\it Publ. RIMS, Kyoto Univ.} \textbf{42}, (2006), 897--922.

\bibitem[HHS05]{hhs}
M. Hirokawa, F. Hiroshima and H. Spohn,
Ground state for point particles interacting through a massless scalar Bose field,
{\it Adv. Math.} {\bf 191} (2005), 339--392.

\bibitem[Hir07]{hir07}
F. Hiroshima, Fiber Hamiltonians in nonrelativistic quantum
electrodynamics, {\it J. Funct. Anal.} {\bf 252} (2007), 314--355.


\bibitem[HIL09]{hil}
F. Hiroshima, T. Ichinose and J. L\H{o}rinczi,
Path integral representation for  Schr\"odinger operators with Bernstein functions of the
Laplacian, arXiv:0906.0103, under review, 2009.

\bibitem[Lov07]{Lov07}
L. Lov\'asz,
\emph{Combinatorial Problems and Exercises}, AMS Chelsea, 2nd edition, 2007.

\bibitem[LHB11]{LHB11}
J. L\H{o}rinczi, F. Hiroshima, and V. Betz,
\emph{Feynman-Kac-Type Theorems and Gibbs Measures on
Path Space. With Applications to Rigorous Quantum Field Theory}, Walter de Gruyter, 2011.

\bibitem[LMS02]{lms}
J. L\H{o}rinczi, R. Minlos, H. Spohn,
The infrared behavior in Nelson's model of a quantum particle coupled to a massless scalar field,
{\it Ann. Henri Poincar\'e} {\bf 3} (2002), 1--28.

\bibitem[Sas05]{sasa}
I.  Sasaki,
Ground state of the massless Nelson model in a non-Fock representation,
{\it J. Math. Phys.} {\bf 46} (2005), 102107.

\bibitem[Spo98]{Sp}
H. Spohn,
Ground state of quantum particle coupled to a scalar boson field,
{\it Lett. Math. Phys.}, {\bf 44} (1998), 9--16.
\end{thebibliography}
}
\end{document}